\documentclass[conference]{IEEEtran}
\usepackage{amsmath,amssymb,amsfonts}
\usepackage{graphicx}
\usepackage{float}
\usepackage{subfig}
\usepackage{color}
\usepackage{cite}
\usepackage{multirow}
\usepackage{algorithm}  
\usepackage{algorithmicx}  
\usepackage{algpseudocode} 
\usepackage{bm}

\usepackage{stfloats}
\usepackage[top=1.8cm,bottom=2.6cm,left=1.7cm,right=1.7cm]{geometry}
\usepackage{siunitx}
\DeclareSIUnit\bps{bps}
\DeclareSIUnit\Torr{Torr}
\DeclareSIUnit\torr{Torr}
\DeclareSIUnit\sample{Sa}
\newcommand*{\circled}[1]{\lower.7ex\hbox{\tikz\draw (0pt, 0pt)%
  circle (.5em) node {\makebox[1em][c]{\small #1}};}}

\ifCLASSINFOpdf

\else

\fi

\graphicspath{{figures/}}

\begin{document}

\title{220 GHz Urban Microcell Channel Measurement and Characterization on a University Campus}

\author{
\IEEEauthorblockN{Yuanbo~Li\IEEEauthorrefmark{1}, Yiqin~Wang\IEEEauthorrefmark{1}, Yejian~Lyu\IEEEauthorrefmark{1}, Ziming~Yu\IEEEauthorrefmark{2}, and Chong~Han\IEEEauthorrefmark{1}}
\IEEEauthorblockA{\IEEEauthorrefmark{1} Terahertz Wireless Communications (TWC) Laboratory, Shanghai Jiao Tong University, China \\ Email:  \{yuanbo.li,wangyiqin,yejian.lyu,chong.han\}@sjtu.edu.cn\\
\IEEEauthorrefmark{2} Huawei Technologies Co., Ltd., China.
Email: yuziming@huawei.com
}
}

\maketitle

\begin{abstract}
Owning abundant bandwidth resources, the Terahertz (THz) band (0.1-10~THz) is envisioned as a key technology to realize ultra-high-speed communications in 6G and beyond wireless networks. To realize reliable THz communications in urban microcell (UMi) environments, propagation analysis and channel characterization are still insufficient. In this paper, channel measurement campaigns are conducted in a UMi scenario at 220~GHz, using a correlation-based time domain channel sounder. 24 positions are measured along a road on the university campus, with distances ranging from 34~m to 410~m. Based on the measurement results, the spatial consistency and interaction of THz waves to the surrounding environments are analyzed. Moreover, the additional loss due to foliage blockage is calculated and an average value of 16.7~dB is observed. Furthermore, a full portrait of channel characteristics, including path loss, shadow fading, K-factor, delay and angular spreads, as well as cluster parameters, is calculated and analyzed. Specifically, an average K-factor value of 17.5 dB is measured in the line-of-sight (LoS) case, which is nearly two times larger than the extrapolated values from the 3GPP standard, revealing weak multipath effects in the THz band. Additionally, 2.5 clusters on average are observed in the LoS case, around one fifth of what is defined in the 3GPP model, which uncovers the strong sparsity in THz UMi. The results and analysis in this work can offer guidance for system design for future THz UMi networks.
\end{abstract}

\IEEEpeerreviewmaketitle

\section{Introduction}
\par Driven by the development of information technologies in the last several decades, the digital world has greatly evolved to support intelligent and colorful applications. Looking towards 2030, the evolution will continue to support many thrilling prospects, such as digital twin, autonomous driving, etc.~\cite{chen2021terahertz} As a result, the data traffic among the intelligent devices and applications will explosively grow, demanding ultra-high communication data rates (e.g., 1 Terabit per second). In light of this, the Terahertz (THz) band, spanning the frequency band from \SI{0.1}{THz} to \SI{10}{THz}, is envisioned as a key technology due to its abundant, contiguous, and yet unregulated bandwidth resource~\cite{Akyildiz2022Terahertz}. 
\par However, challenges remain to realize effective THz communications, where a fundamental one lies in the full understanding of the THz channels, which requires extensive measurement efforts. During the last decade, the THz channel sounder systems have been greatly developed. Therefore, recently a few research groups have built up THz channel sounders and conducted channel measurement campaigns in various scenarios and frequency bands~\cite{Eckhardt2021channel,ju2023ghz,Abbasi2023Band,Abbasi2023THz,chen2021channel,li2023channel,wang2023thz,li2024pico}, as summarized in~\cite{han2022terahertz}. Among many scenarios, the urban microcell (UMi) is a key use case for THz communications to support outdoor coverage. Though a few works have been done to investigate the THz channels in UMi~\cite{Abbasi2023Band,Xing2021Propagation,Lee2023subthz}, the investigated frequency bands are limited to \SI{140}{GHz} or \SI{159}{GHz} band and the measurement distance are only around \SI{100}{m}. Moreover, analysis of the interaction of the THz wave and surrounding environments in the THz UMi, especially the foliage blockage effects, is still insufficient.
\par In this paper, based on a correlation-based time-domain channel sounder, channel measurement campaigns are conducted on a UMi scenario at \SI{220}{GHz}. Overall, 24 receiver (Rx) positions are measured, where the distance between the transmitter (Tx) and Rx ranges from \SI{34}{m} to \SI{410}{m}. Based on the measurement results, spatial consistency analysis is conducted by observing the evolution of multipath components (MPCs) along the Rx route, based on which the interaction of THz waves and surrounding environments is studied. Moreover, the additional loss due to foliage blockage is studied and statistically modeled. Last but not least, the channel characteristics, including path loss, shadow fading, K-factor, delay and angular spreads, as well as cluster parameters, are calculated and analyzed. The channel characterization results are compared with existing 3GPP channel standards.
\par The remainder of the paper is organized as follows. In Sec.~\ref{sec:measurement}, the measurement system, setup and deployment are demonstrated. Furthermore, the channel measurement results are analyzed in Sec.~\ref{sec:char}. Finally, Sec.~\ref{sec:conclude} concludes the paper. 
\section{Channel Measurement Campaign}
\label{sec:measurement}
\subsection{Measurement System}
\label{sec:system}
\begin{figure}
    \centering
    \subfloat[Diagram of the channel sounder.] {
     \label{fig:sounder}     
    \includegraphics[width=0.9\columnwidth]{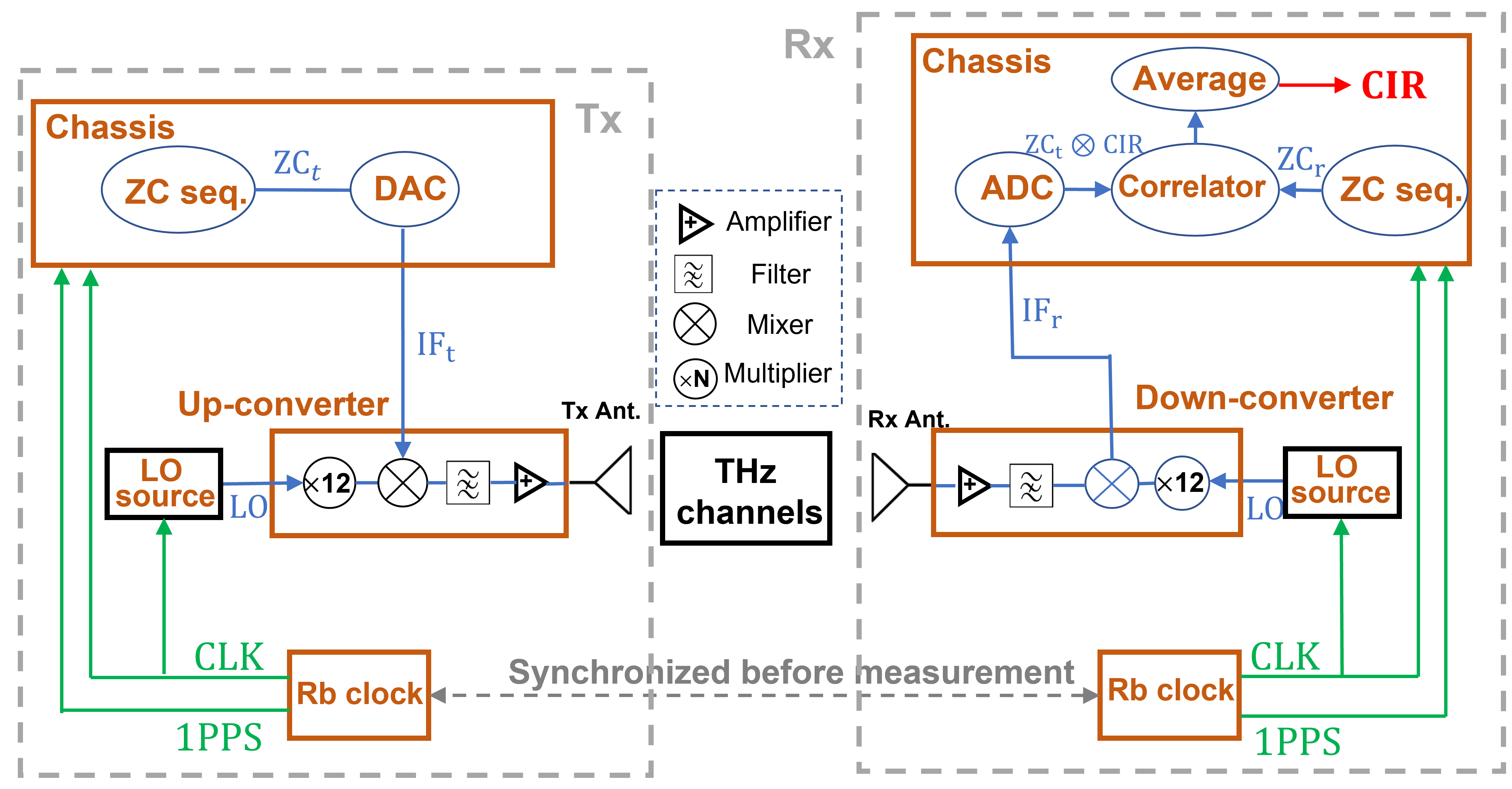}  
    }
    \quad
    \subfloat[Picture of hardware.] {
     \label{fig:hardware1}     
    \includegraphics[width=0.9\columnwidth]{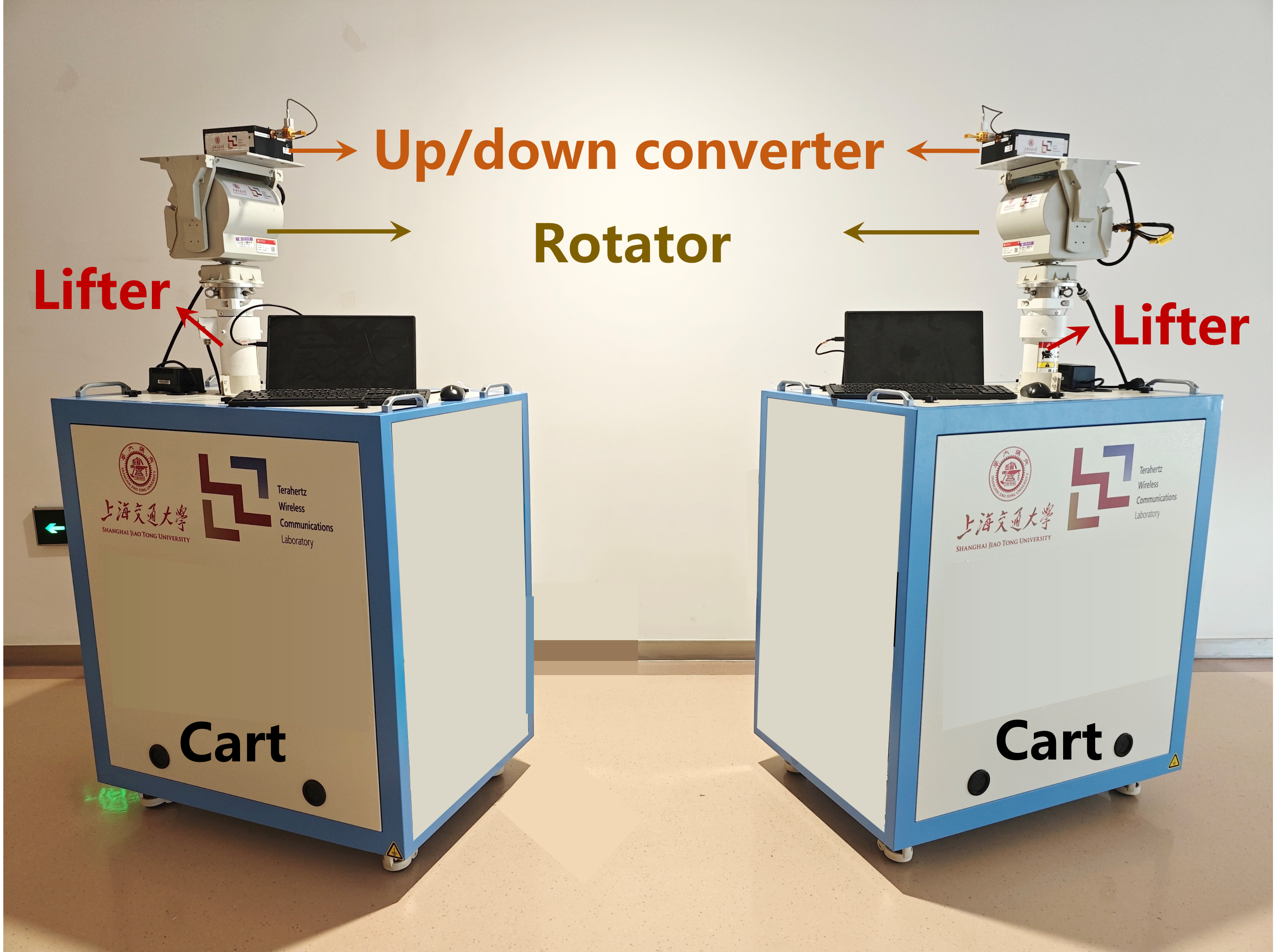}  
    }
    \caption{Correlation-based time-domain channel sounder.}
    \label{fig:cs}
    \vspace{-0.5cm}
\end{figure}
\par To measure THz channels, a correlation-based time-domain channel sounder is used, as shown in Fig.~\ref{fig:cs}. The channel sounder system consists of a Tx subsystem and a Rx subsystem, both of which include a baseband processing chassis, a radio frequency front end, and a mechanical platform for the movement of Tx/Rx. Two rubidium (Rb) clocks are utilized to separately provide clock reference and 1 pulse-per-second (1PPS) trigger to Tx and Rx subsystems, which are connected in a master-slave mode for several hours before real measurements to synchronize the 1PPS signal. For a detailed description of the channel sounder, the readers may refer to our previous work in~\cite{Li2023Correlation}. 
\par It is noteworthy that no cable connection is needed when using the correlation-based sounder, for which it is much more convenient to conduct long-distance outdoor channel measurements than the vector network analyzer (VNA)-based channel sounder~\cite{chen2021channel,li2023channel,wang2023thz}. Moreover, through a 5000-times average, the noise floor of the measured channel impulse responses (CIRs) is lower than \SI{-170}{dB}, which supports multipath measurements over \SI{400}{m} Tx-Rx separation distance, as verified in this work. 
\subsection{Measurement Campaign}
\par The measurement campaign is conducted on Nanyangdong Road in the Shanghai Jiao Tong University campus. There are several buildings on both sides of the road, such as the Longbin Building, Wenbo Building, and buildings of the School of Mechanical Engineering (ME), etc. The heights of the surrounding buildings are \SIrange{10}{20}{m}. Evergreen trees are densely planted along the road, with heights of around \SI{5}{m}.
\begin{figure}[!tbp]
    \centering
    \includegraphics[width=0.9\columnwidth]{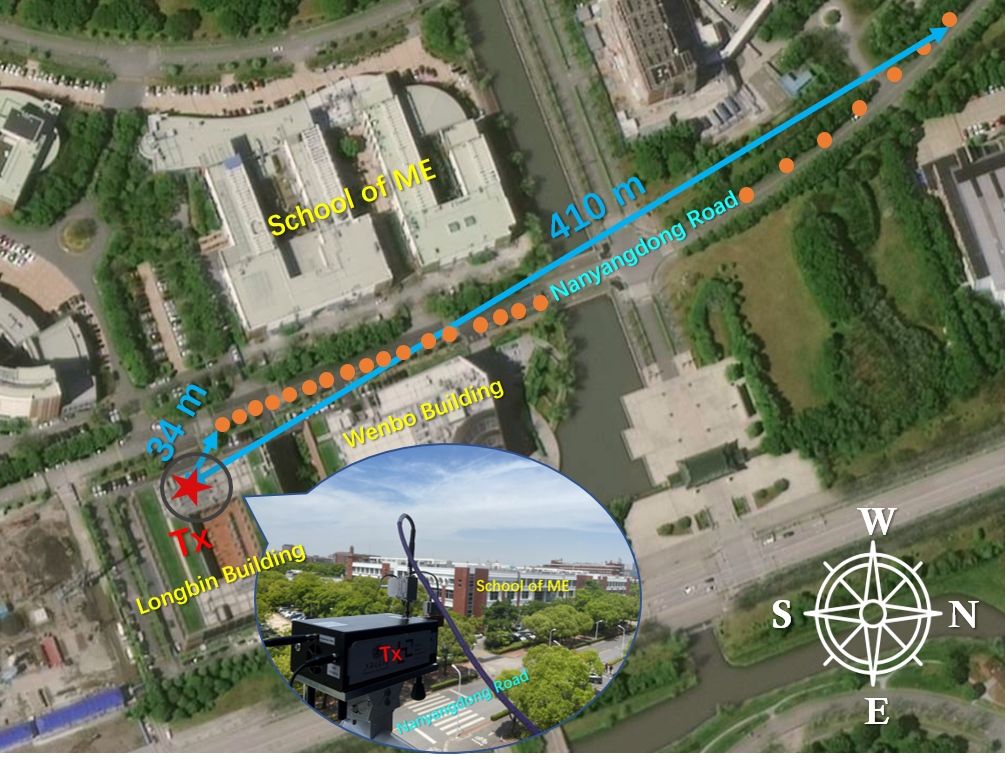}  
    \caption{The measurement deployment in the UMi scenario. Rx positions are labeled with orange dots.}
    \label{fig:layout}
    \vspace{-0.5cm}
\end{figure}
\par To imitate a UMi base station, the Tx is deployed on a horizontal pole and reaches out of a window on the fourth floor in the Longbing Building, achieving a \SI{16.6}{m} height above the ground, as shown in the inserted picture in Fig.~\ref{fig:layout}. Additionally, Rx imitates user equipment with a height of \SI{1.6}{m}, distributed along the Nanyangdong Road, as labeled with orange dots in Fig.~\ref{fig:layout}. Overall, 24 Rx positions are measured, where 9 of them have clear line-of-sight (LoS) propagation and the rest are partially obstructed by the foliage alongside the road, leading to the obstructed-LoS (OLoS) case. The Tx-Rx distance ranges from \SI{34}{m} to \SI{410}{m}. 

\subsection{Measurement Setup}
\par The measured frequency band is centered at \SI{220}{GHz}, with a bandwidth of \SI{1.536}{GHz}. As a result, the delay resolution is \SI{0.65}{ns}, which indicates that any two multipath components (MPCs) with propagation distance difference larger than \SI{19.5}{cm} can be distinguished. Moreover, the measured CIRs contain 2048 temporal samples, resulting in a maximum delay of \SI{1332.7}{ns} and a maximum path length of \SI{399.8}{m}. 
\par Note that at the farthest Rx position, the Tx/Rx distance is slightly larger than the maximum path length of our sounder, resulting in circular shifted CIR results. To address this, the CIRs measured at the last Rx position are manually processed, i.e., the CIR sequences are extended to a length of 2154 samples, with the last 106 samples copied from the first 106 samples in the original CIR results, resulting in an equivalent \SI{420}{m} measurable path length.
\par Moreover, the Tx directly radiates out THz waves through a standard waveguide WR5, which has a \SI{7}{dBi} antenna gain. By contrast, the Rx is equipped with a horn antenna with a \SI{26}{dBi} gain and a $8^\circ$ half-power beamwidth (HPBW). To capture MPCs from various directions, direction-scan sounding (DSS) is conducted through mechanical rotation of the Rx, scanning the azimuth plane from $0^\circ$ to $360^\circ$ and elevation plane $-20^\circ$ to $20^\circ$ with a $10^\circ$ angle step. 
\begin{figure*}[!tbp]
    \centering
    \subfloat[PDPs.]{
    \centering
    \includegraphics[width=1.0\columnwidth]{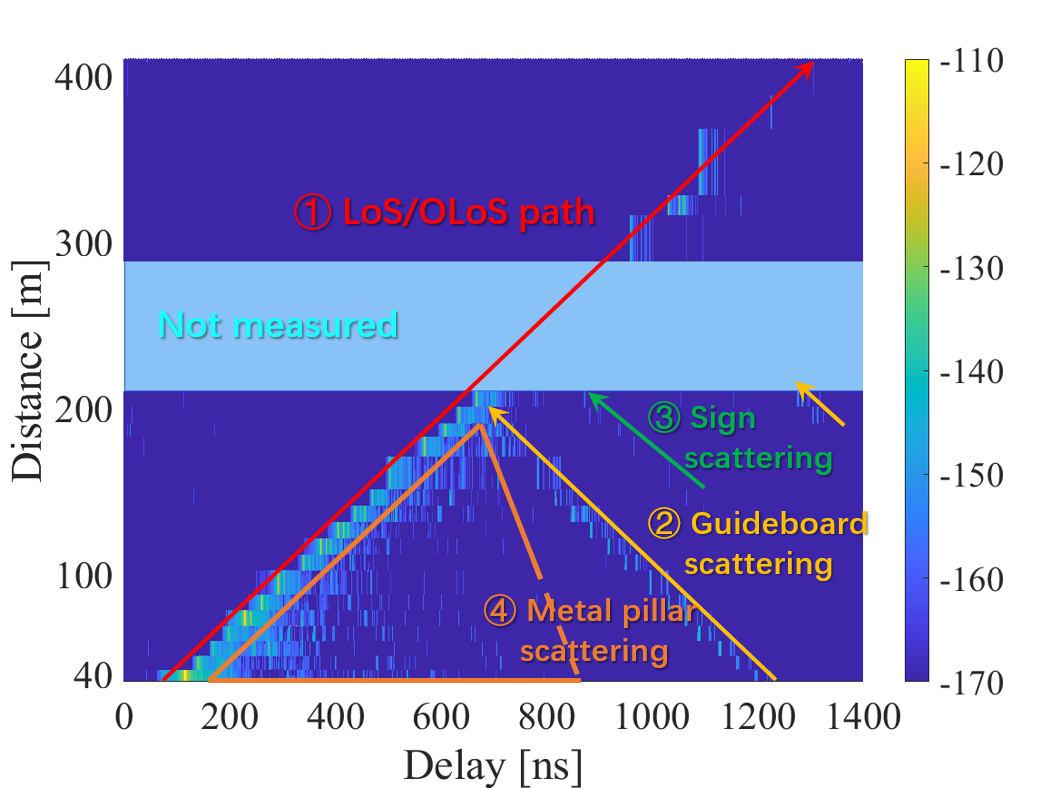}  
    }
    \subfloat[Objects.]{ 
    \centering
    \includegraphics[width=0.9\columnwidth]{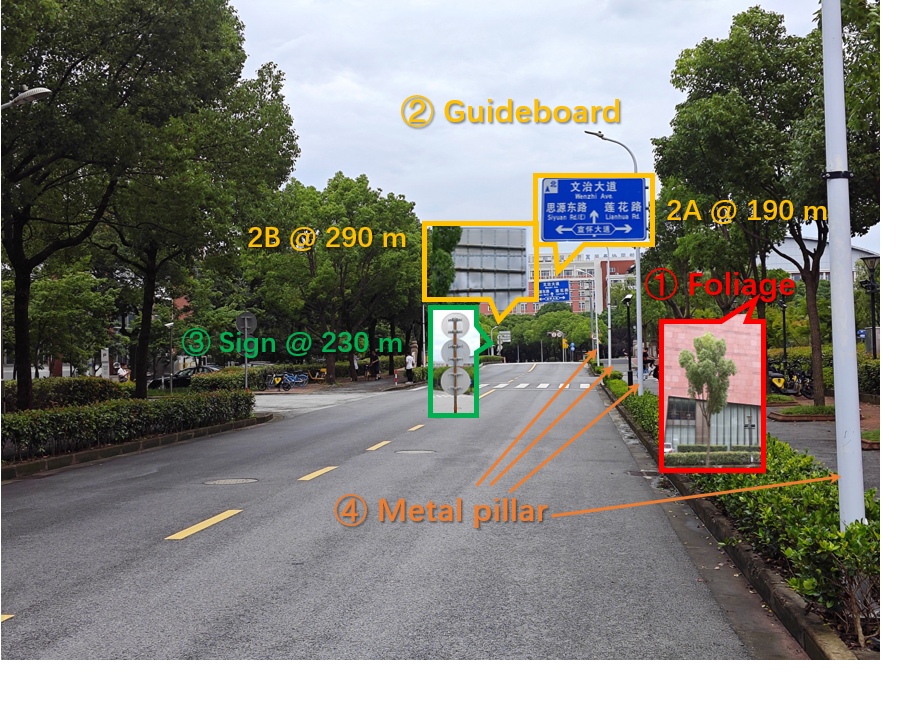}  
    }
    \caption{The spatial consistency analysis along the Nanyangdong Road. PDPs measured in different Rx locations are shown in (a), while the objects affecting the THz propagation are shown in (b).}
    \label{fig:sp}
    \vspace{-0.5cm}
\end{figure*}
\section{Measurement Results and Analysis}
\label{sec:char}
\subsection{Data Post-processing}
\par The measured raw CIR data are processed through calibration, time-drift correction, channel estimation, and MPC clustering. First, to eliminate the system response, a back-to-back calibration process is implemented. Second, though the two Rb clocks of Tx and Rx subsystems are synchronized before the real measurements, the two 1PPS signals from them will gradually drift apart as time goes by. To address this problem, a linear interpolation/extrapolation method is used, as detailed in our previous work~\cite{Li2023Correlation}. Third, as the spatial profiles are measured through DSS, the antenna response affects the measured results, which is further eliminated through accurate channel parameter estimation by using the DSS-o-SAGE algorithm~\cite{li2024sage}. Last but not least, to observe the cluster parameters in THz channels, the MPCs are further divided into clusters by using a multipath component distance (MCD)-based density-based spatial clustering of applications with noise (DBSCAN) algorithm, which shows good clustering performance in our previous work~\cite{chen2021channel}. Note that to save processing time, during the channel parameter estimation process, only the MPCs with path gains larger than a certain threshold are extracted, where the path gain threshold is set as
\begin{equation}
    \alpha_\text{th} = 10^{(\max{(\alpha_1-\text{R},\text{NF}+5)}/20)},
\end{equation}
where \text{R} denotes the dynamic range, which is set as \SI{30}{dB} in this work. Moreover, $\text{NF}$ stands for the noise floor of the measured CIRs, which is \SI{-170}{dB} for our sounder. Additionally, $\max{(\text{A},\text{B})}$ produces the maximum value between A and B.
\subsection{Propagation Analysis}
\label{sec:scandpa}
\par As the locations of the Rx change, variations of the channel status can be observed. However, as closely-placed Rxs share similar propagation environments, their channels are not independent but spatially correlated, which is termed as the \textit{spatial consistency}. To study this, the power delay profiles (PDPs) in different Rx locations are connected together and shown in Fig.~\ref{fig:sp}. Note that the PDPs at each Rx location are produced by summing the noise-eliminated PDPs measured in all directions. The noise samples in PDPs measured in each direction, namely those samples with power smaller than \text{NF}, are set to a nearly zero value, e.g., -200 dB, to avoid the rise of the noise floor due to superimposition of multiple PDPs.
\par Several observations can be made as follows. First, a clear trend that can be observed is the evolving of the LoS/OLoS path along the movement of the Rx. As the propagation delay of the LoS/OLoS path is positively correlated to the Tx-Rx distance, the trajectory of the LoS/OLoS path is a straight line with a slope of the speed of light. Moreover, the signal strength of the LoS/OLoS path constantly fluctuates due to the occasional blockage by foliage. At the farthest Rx position with \SI{410}{m} distance, only the OLoS path can be observed with a weak path gain of \SI{-160}{dB}. Second, another obvious trajectory originates from the scattering path from the guideboard that is around \SI{190}{m} away from the Tx. As the Tx-Rx distance increases, the propagation delay of the guideboard-scattered path decreases, which eventually vanishes as the Rx moves farther away than the guideboard. There is another guideboard at around \SI{290}{m} distance from the Tx, denoted as 2B in Fig.~\ref{fig:sp}(b), where the scattered path can be slightly seen when the distance is close to \SI{200}{m}. However, the scattered signal is very weak since it stems from the back of the guideboard 2B. Third, similar to the guideboard scattering, within the distance range \SIrange{140}{200}{m}, the scattered path from a sign with \SI{230}{m} distance away from the Tx can be observed, whose signal strength is also weak. Fourth, the metal pillars on both sides of the street, such as those of lampposts, signs, etc., cause rich scattering along the Rx route. However, the signal strength is weak and spatial consistency is not obvious since the pillars are too small to affect a large area. 
\subsection{Foliage Blockage and Attenuation}
\begin{figure}
    \centering
    \subfloat[Propagation loss of LoS/OLoS path.] {  
    \includegraphics[width=0.85\columnwidth]{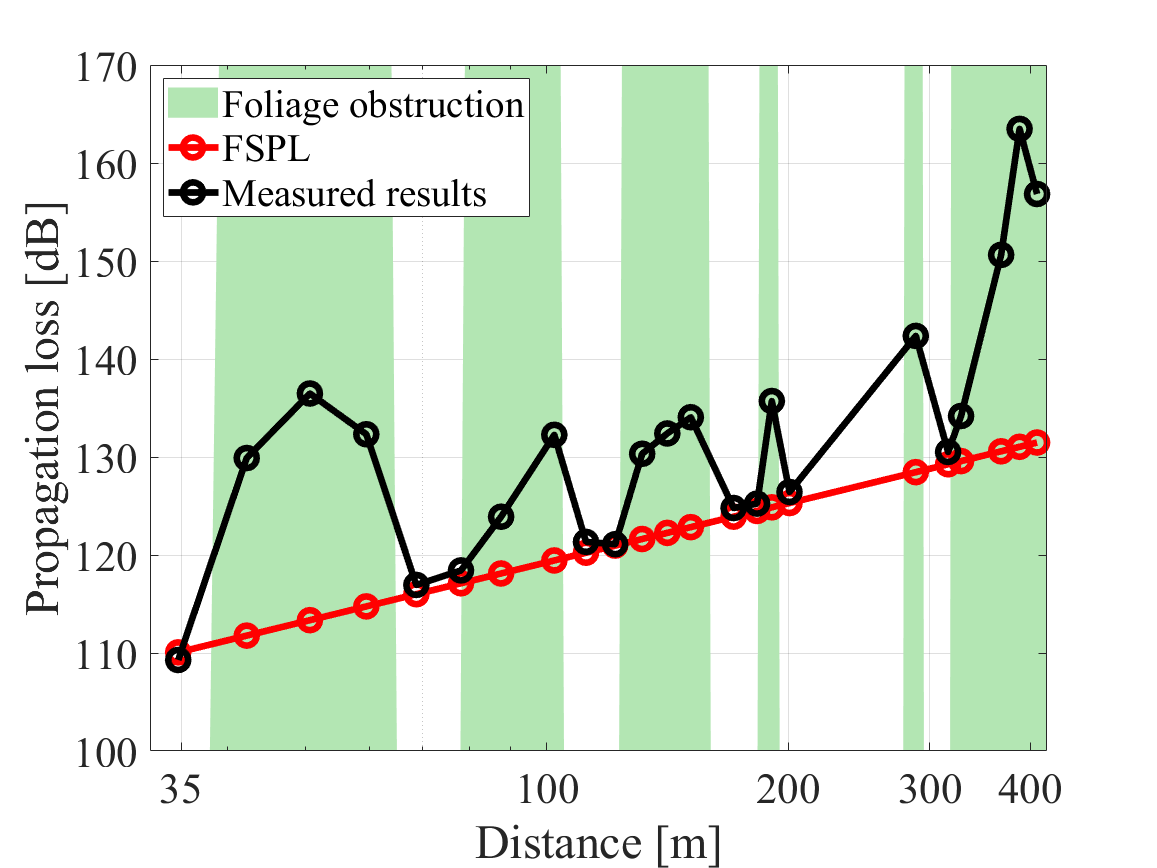}  
    }
    \quad
    \subfloat[Foliage loss.] { 
    \includegraphics[width=0.85\columnwidth]{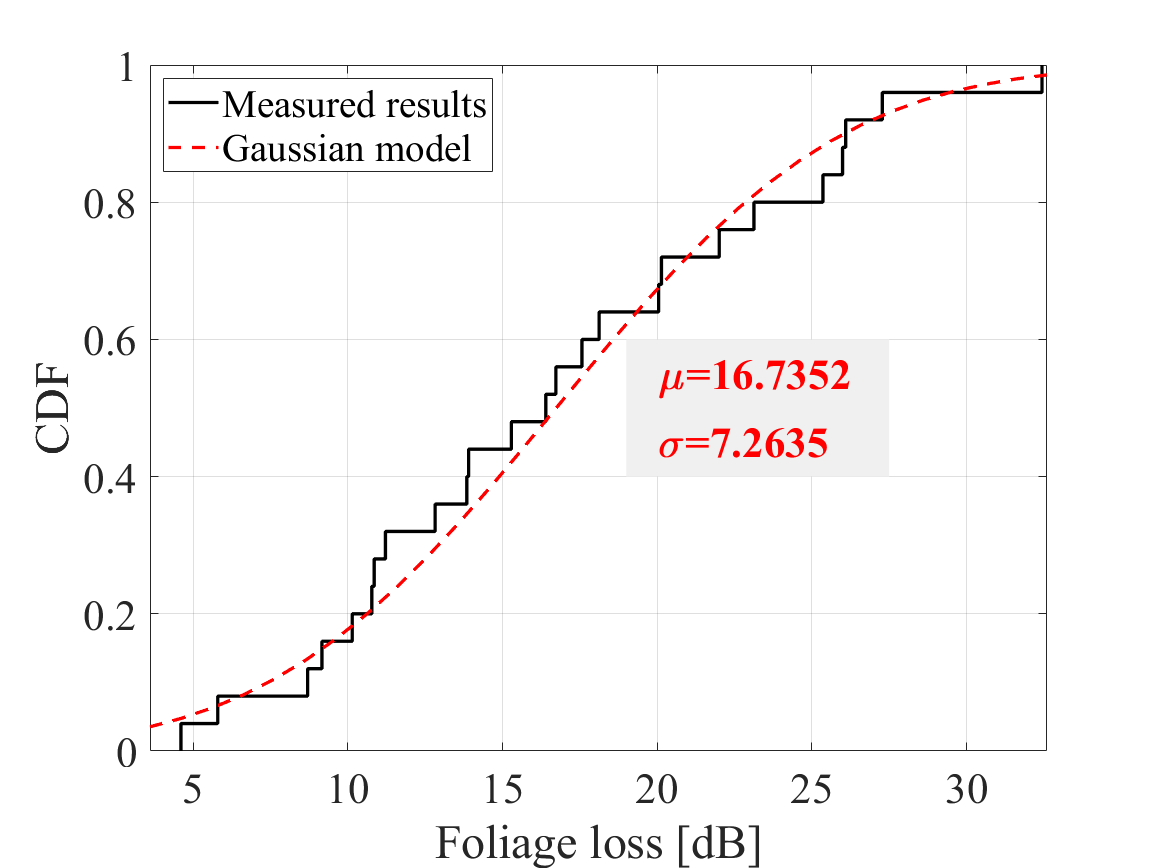}  
    }
    \caption{The propagation loss of the LoS/OLoS path and additional loss due to foliage.}
    \label{fig:foliage}
    \vspace{-0.5cm}
\end{figure}
\par As elaborated in the propagation analysis, the foliage alongside the road causes occasionally blockage to the LoS path, resulting in signal strength fluctuations of the LoS/OLoS path, as shown in Fig.~\ref{fig:foliage}(a). The propagation loss of the LoS/OLoS path is calculated relating the path gain, i.e., $\text{PL}_\text{LoS/OLoS}=-20\log_{10}(\alpha_{\text{LoS/OLoS}})$ with $\alpha_{\text{LoS/OLoS}}$ denoting the path gain of the LoS/OLoS path. It can be seen that foliage obstruction occasionally occurs along the Rx route. Due to different foliage thicknesses, the additional loss from the foliage obstruction ranges from several dB to tens of dB.
\par To fully study the foliage blockage, the foliage loss is further calculated by subtracting the free space path loss (FSPL) from the propagation loss of the LoS/OLoS path, namely
\begin{equation}
    \text{L}_\text{Fol} [\text{dB}]=\text{PL}_\text{LoS/OLoS}-\text{FSPL}(f_0,c\tau_{\text{LoS/OLoS}})
\end{equation}
where $\tau_{\text{LoS/OLoS}}$ denotes the delay of the LoS/OLoS path. Moreover, $\text{FSPL}(f,d)$ represents the FSPL at frequency $f$ and distance $d$, which can be calculated as $\text{FSPL}(f,d)=20\log_{10}(c/4\pi fd)$. Additionally, $c$ is the speed of light.
\par The cumulative distribution function (CDF) of the foliage loss is drawn in Fig.~\ref{fig:foliage}(b), which is fitted using a Gaussian distribution. The foliage loss in the THz band ranges from \SI{5}{dB} to \SI{32}{dB}, depending on foliage thickness. Moreover, the Gaussian fitting shows good performance, with a mean of \SI{16.74}{dB} and a standard deviation of 7.26.
\subsection{Statistical Modeling of Channel Characteristics} 
\begin{table}[tbp]
    \centering
    \caption{Measured channel characteristics in the THz UMi.}
    \includegraphics[width = 0.75\columnwidth]{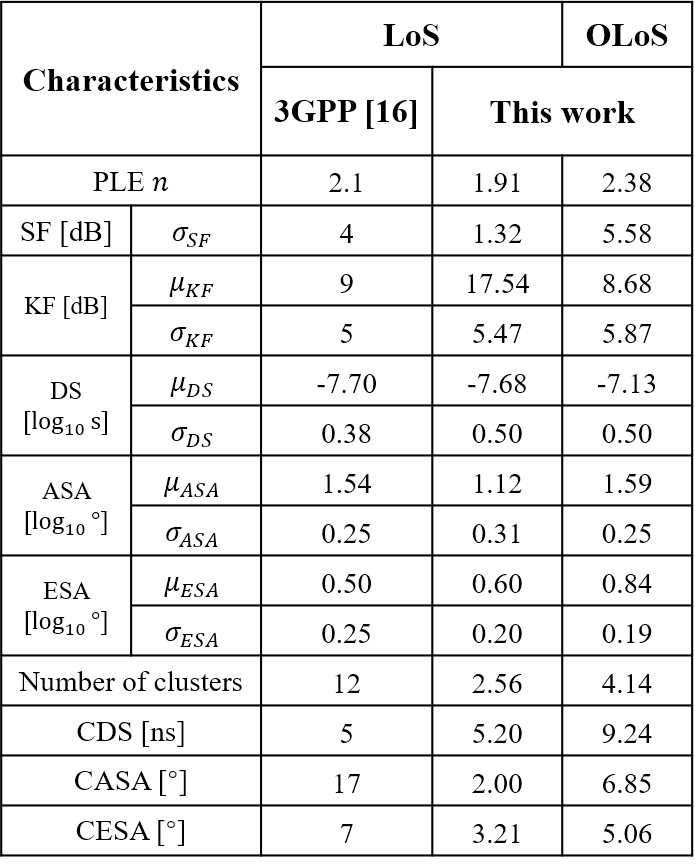}
    \label{tab:char}
    \vspace{-0.5cm}
\end{table}
\par Following the framework of 3GPP standardization files~\cite{3gpp.38.901}, the key channel characteristics are calculated and fitted with statistical models. The fitting parameters are listed in Table~\ref{tab:char}, which are analyzed as follows.
\subsubsection{Path loss and shadow fading}
\par Known as large-scale fading, path loss and shadow fading are key parameters for link budget analysis. The path loss results are fitted using the close-in (CI) free space reference distance model, while the shadow fading is termed as a zero-mean Gaussian distributed random variable. First, the measured path loss exponents (PLE) are 1.91 and 2.38 in LoS and OLoS cases, respectively. The PLE in the LoS case is close to 2, namely the PLE in free space, indicating that the LoS path contributes most of the received power. The slightly larger PLE in OLoS cases compared to LoS cases is attributed to the foliage blockage. Second, weak shadow fading effects are observed in the LoS case, with a standard deviation of \SI{1.32}{dB}. In contrast, the shadowing effect is more significant in the OLoS case, where the standard deviation of the shadow fading is \SI{5.58}{dB}. 
\subsubsection{K-factor}
\par The K-factor evaluates how dominant the strongest cluster is. The K-factor results are fitted following log-normal distribution, as shown in Table~\ref{tab:char}. First, the mean K-factor values are \SI{17.54}{dB} and \SI{8.68}{dB} in LoS and OLoS cases, respectively. The large K-factor values in the LoS case indicate the strong dominance of the LoS cluster. In the OLoS case, the THz channels are still dominated by one strong cluster, mostly the OLoS cluster. 
\subsubsection{Delay and angular Spreads}
\par The power of MPCs disperses in both temporal and spatial domains, which therefore affects the communication system designs. The power dispersion can be characterized by delay and angular spreads. The delay spreads (DS), azimuth spreads of arrival (ASA) and elevation spreads of arrival (ESA) are calculated and fitted with log-normal distributions, as shown in Table~\ref{tab:char}. First, the average delay spread values are \SI{20.89}{ns} and \SI{74.13}{ns} in the LoS and OLoS cases, respectively. The delay spreads are much larger than the reported values in our previous work in THz picocell scenarios, where the DS values are mostly several nanoseconds in the LoS case and up to \SI{20}{ns} in the NLoS case~\cite{wang2023thz,li2024pico}. The reason is that the base station height in this work is much higher than those in our previous work, for which MPCs with very large delays can also be received, such as the scattering paths from guideboards and signs, resulting in large DS values. 
\par Second, the average ASA values are $13.18^\circ$ and $38.90^\circ$ for LoS and OLoS cases, respectively, while the mean ESA values are $3.98^\circ$ and $6.92^\circ$ for LoS and OLoS cases, respectively. The results indicate stronger multipath effects in the OLoS case, since the OLoS path is weaker than the LoS path and the effects of other NLoS paths are more significant.
\subsubsection{Cluster parameters}
\par The cluster parameters are calculated, including the number of clusters, cluster delay spread (CDS), cluster azimuth spread of arrival (CASA) and cluster elevation spread of arrival (CESA), whose values are discussed as follows. First, the average CDS, CASA, and CESA values are smaller in the LoS case, which is consistent with the observations for delay and angular spreads. Second, the average numbers of clusters are 2.56 and 4.14 in the LoS and OLoS cases, indicating the strong sparsity in the THz UMi scenarios.
\subsubsection{Comparison with 3GPP model}
\par The reference values in 3GPP TR 38.901 for the UMi scenario are used for comparison~\cite{3gpp.38.901}. Note that as the OLoS case, where the LoS path is partially blocked by objects, is not defined as a separate case in the 3GPP model, only the measured results in the LoS case is compared with the reference values in the 3GPP model. Moreover, it is noteworthy that as the investigated frequency band is beyond the applicability of the 3GPP model, the parameters extrapolated from the 3GPP model only show what one might expect in the THz band. 
\par Several observations can be made as follows. First, the 3GPP model overestimates the large-scale fading parameters, namely the path loss and shadow fading terms. Second, a giant difference occurs for the K-factor values, where much larger K-factors are measured than that one may expect by using the 3GPP model. Third, the delay and angular spreads are close, except that the ASA values are measured to be slightly smaller than the results in the 3GPP model. Fourth, the measured number of clusters is much smaller than the reported values in the 3GPP standard, indicating the significant sparsity in the THz channels. To summarize, compared to the expected channel characteristics obtained using the 3GPP standard, the real measured results reveal that the THz channels exhibit slightly smaller large-scale fading values, much larger K-factor values, weaker multipath effects, and strong sparsity.
\section{Conclusion}
\label{sec:conclude}
In this paper, we conducted measurement campaigns on an UMi scenario at \SI{220}{GHz} using a correlation-based channel sounder. Along a street on the university campus, 24 Rx positions are measured, with Tx-Rx distance ranging from \SI{34}{m} to \SI{410}{m}. Based on the measurement results, we analyzed the spatial consistency of MPCs and the interactions of the THz wave and surrounding environments. We calculated the additional loss due to foliage blockage and reported an average value of \SI{16.7}{dB}. Furthermore, we calculated and analyzed the channel characteristics, including path loss, shadow fading, K-factor, delay and angular spreads, as well as cluster parameters. Specifically, in the LoS case, an average K-factor value of 17.5 dB is measured, nearly two times larger than the extrapolated values from the 3GPP standard, which reveals the weak multipath effects in the THz band. Additionally, 2.5 clusters on average are observed in the LoS case, which is around one fifth of the values specified in the 3GPP model, uncovering the strong sparsity in THz UMi.
The measurement results and analysis in this work are helpful for the system design of future THz UMi networks.
\bibliographystyle{IEEEtran}
\bibliography{IEEEabrv,main}

\end{document}